\documentclass[pra,showpacs]{revtex4}

\usepackage[intlimits]{amsmath}
\usepackage{graphicx}

\def\ket#1{|#1\rangle}
\def\bra#1{\langle#1|}
\def\inner#1#2{\langle #1 | #2 \rangle}
\def\msum#1{\sum_{#1=-\infty}^{\infty}}
\def\Dsum#1{\sum_{#1=0}^{D-1}}

\begin{document}

\newcommand{\re}{\,\text{Re}\,}
\newcommand{\im}{\,\text{Im}\,}

\title{The classical limit for a class of quantum baker's maps}

\author{Mark M. Tracy}
\email{mtracy@phys.unm.edu}
\author{A. J. Scott}
\email{ascott@phys.unm.edu}
\affiliation{Department of Physics and Astronomy, University of New Mexico,  
Albuquerque, New Mexico 87131-1156, USA }

\date{1 June 2002}

\begin{abstract}
We show that the class of quantum baker's maps defined by Schack and Caves have 
the proper classical limit provided the number of momentum bits approaches infinity. 
This is done by deriving a semi-classical approximation to the coherent-state 
propagator. 
\end{abstract}


\maketitle

\section{Introduction}

The introduction of `toy' mappings which demonstrate essential features of nonlinear 
dynamics has led to many insights in the field of classical chaos. A well-known example 
is the so-called {\it baker's transformation} \cite{Lichtenberg1992a}. Interest in 
this mapping stems from its straightforward characterization in terms of a Bernoulli 
shift on binary sequences. It seems natural to consider a quantum counterpart to the 
baker's map for the investigation of quantum chaos. Unfortunately, there is no unique 
quantization procedure, and hence, we must embrace the possibility 
of different quantum maps limiting to the same classical baker's transformation. 

Balazs and Voros \cite{Balazs1989a} were first to conceive a quantum version of the 
baker's map. This was done with the help of the discrete quantum Fourier transform. 
Subsequently, improvements to the Balazs-Voros quantization were made by Saraceno 
\cite{Saraceno1990a}, an optical analogy was found \cite{hannay}, a canonical 
quantization was devised \cite{Rubin1998a,lesniewski}, and quantum computing realizations have 
been proposed \cite{schack,brun}. A quantum baker's mapping on the sphere has also been defined 
\cite{Pakonski1999a}. More recently, an entire class of quantum baker's maps was 
proposed by Schack and Caves using qubits \cite{Schack2000a}. The Balazs-Voros 
quantization is but one member of this class.

The classical limit of the Schack-Caves quantizations is the subject of this article. 
We explicitly derive a semi-classical approximation for the propagator in the coherent 
state basis. This enables us to give conditions upon which the Schack-Caves quantizations 
will behave as the classical baker's transformation in the limit $\hbar\rightarrow 0$.
We find that, provided the number of momentum qubits approaches infinity, the 
semi-classical propagator takes the form 
\[
\langle b|\hat{B}| a\rangle \approx \sqrt{\frac{\partial^2 W}{\partial a\partial b^*}}
\exp\Big[W(b^*,a)/2\hbar\Big]\exp\Big[-\big(|a|^2+|b|^2\big)/4 \hbar\Big]
\]
where $|a\rangle$ and $|b\rangle$ are coherent states on the torus, and $W(b^*,a)$ is a 
classical generating function. Similar propagators have been encountered before using spin coherent 
states \cite{Kus1993a,Scott2001a}, but all may be thought of as variants of those derived long 
ago by Van Vleck \cite{Van1928a} and Gutzwiller \cite{Gutzwiller1967a}. Semi-classical 
propagators play an important role in the path-integral 
formulation of quantum mechanics \cite{Feynman} and the related theory of periodic orbit 
quantization \cite{Brack}. The latter has been investigated thoroughly for the Balazs-Voros 
quantum baker's map \cite{ozorio,eckhardt,saraceno,dittes,laksh,luz,kaplan,toscano,tanner}.

In deriving a semi-classical approximation only for the one-step propagator, 
we avoid complications which will arise after many iterations of the mapping.
For long time scales, simple quantum-to-classical correspondences will break down \cite{inoue} and one 
must incorporate the theories of decoherence \cite{Zurek,Giulini,Habib} or continuous measurement 
\cite{Bhatt,Scott2000a}. The classical limit of the Schack-Caves quantization has  
already been investigated \cite{Soklakov2000a} in this light using a decoherent histories
approach \cite{Griffiths1984a,Omnes1988a,Gell1993a}. However only a special case ($\theta=0$ in 
our notation) was considered. We proceed under an assumption that provided our one-step 
propagator agrees with the baker's transformation in the semi-classical limit $\hbar\rightarrow 0$, 
decoherence will restore quantum-to-classical correspondences for long time scales. 

The paper is organized as follows. In Section II, we introduce the baker's map, both
in classical and quantal form. Coherent states for a toroidal phase space are also 
introduced. In Section III our core results are presented. Here we derive 
semi-classical approximations to the coherent-state propagator and give conditions 
for when the Schack-Caves quantizations have the proper classical limit. Finally, 
in Section IV, we summarize our findings.

\section{A Class of Quantum Baker's Maps}

The baker's map is a standard example in chaotic
dynamics. It is a mapping of the unit square onto itself in the form
\begin{eqnarray}
q_{n+1} &=& 2 q_n - \lfloor 2q_n \rfloor \label{classbake1}\\
p_{n+1} &=& \left(p_n + \lfloor2 q_n \rfloor \right)/2 \label{classbake2}
\end{eqnarray}
where $q,p \in [0,1)$, $\lfloor x \rfloor$ is the integer part of $x$, and $n$
denotes the $n$-th iteration of the map. Geometrically,
the map stretches the unit square by a factor of two in the $q$ direction, squeezes
by a factor of a half in the $p$ direction, and then stacks the right half onto the left.

The map's action may be rewritten in terms of the complex variable $a\equiv q+ip$,
\begin{equation}
a_{n+1}^{\phantom{*}} = \frac{5}{4} a_n^{\phantom{*}} + \frac{3}{4} a_n^* + \left(\frac{i}{2}-1\right) \lfloor a_n^{\phantom{*}}+a_n^*
\rfloor \equiv b_n^{\phantom{*}}(a_n^{\phantom{*}},a_n^*).\label{classbake3}
\end{equation}
A generating function for this mapping (up to an arbitrary constant) is 
\begin{equation}
\label{genfunc}
W(b^*,a) = \frac{1}{10}\left(3b^{*2}+8ab^*-3a^2\right) + \frac{4}{5}\left(1+\frac{i}{2}\right)
\left(a+ib^*-\frac{1}{2}\right) \lfloor a+a^* \rfloor
\end{equation}
assuming $a+a^*$ is non-integer. The classical baker's map may then be rederived {\it via}
the relations
\begin{equation}
\frac{\partial W}{\partial b^*} = b \qquad\qquad
\frac{\partial W}{\partial a} = a^*.
\end{equation}

Interest in the baker's map is due mainly to the simplicity of its {\it symbolic dynamics}.
If each point of the unit square is identified through its binary representation, 
$q = 0\cdot s_1 s_2 \ldots = \sum_{k=1}^\infty s_k 2^{-k}$ and
$p= 0\cdot s_0 s_{-1} \ldots = \sum_{k=0}^\infty s_{-k} 2^{-k-1}$ ($s_i\in\{0,1\}$), with a bi-infinite symbolic string
\begin{equation}
s = \ldots s_{-2} s_{-1} s_0 \bullet s_1 s_2 s_3 \ldots \label{symbseq}
\end{equation}
then the action of the baker's map is to shift the position of the dot by one point to the
right,
\begin{equation}
s\rightarrow s' = \ldots s_{-2} s_{-1} s_0 s_1 \bullet s_2 s_3 \ldots .
\end{equation}

For a quantum mechanical version of the map, we work in the $D$-dimensional Hilbert space,
${\cal H}_D$, spanned by either the position states $\ket{q_j}$, with eigenvalues 
$q_j =(j+\beta)/D$, or momentum states $\ket{p_k}$, with eigenvalues $p_k =(k+\alpha)/D$ 
($j,k=0 \ldots D-1$). The constants $\alpha,\beta\in [0,1)$ determine the periodicity of the space:
$\ket{q_{j+D}} = e^{-2 \pi i \alpha} \ket{q_j}$, $\ket{p_{k+D}} = e^{2 \pi i \beta} \ket{p_k}$.
Such double periodicity identifies ${\cal H}_D$ with a toroidal phase space.
The vectors of each basis are orthonormal 
$\inner{q_j}{q_{j'}} = \delta_{j,j'}$, $\inner{p_k}{p_{k'}} = \delta_{k,k'}$ and the two bases are
related {\it via} the finite Fourier transform
$$ \bra{q_j}\hat{F}_D\ket{q_k}\equiv\inner{q_j}{p_k} = \frac{1}{\sqrt{D}} e^{\frac{i}{\hbar} q_j p_k} .$$
For consistency of units, we must have $2 \pi \hbar D=1$.

The first work on a quantum baker's map was done by Balazs and Voros \cite{Balazs1989a}.  Their
expression for the map was given in the form
\begin{equation}
\hat{B} = \hat{F}^{-1}_D \left(
\begin{array}{cc}
\hat{F}_{D/2} & 0 \\
0 & \hat{F}_{D/2}
\end{array} \right)
\end{equation}
where $\hat{F}_{D/2}$ is the finite Fourier transform acting on half of the Hilbert space.
Later Saraceno \cite{Saraceno1990a} improved certain symmetry characteristics of the map
using anti-periodic boundary conditions ($\alpha=\beta=1/2$).
Finally, taking again the anti-periodic Hilbert space, Schack and Caves \cite{Schack2000a} introduced a whole class of quantum baker's
maps for dimensions $D=2^N$.

For these cases, we can model our space as the product of $N$ qubits with a binary
expansion association
\begin{equation}
\ket{q_j} = \ket{x_1} \otimes \ket{x_2} \otimes \cdots \otimes \ket{x_N} \qquad x_l\in\{0,1\} \label{qj}
\end{equation}
where $j$ has the binary expansion  
\begin{equation}
j=x_1 \ldots x_N\cdot 0 = \sum_{l=1}^N x_l 2^{N-l}\qquad\text{and}\qquad q_j=\frac{j+1/2}{D}.
\end{equation}
Next we rewrite the quantum Fourier transform as
\begin{equation}
\ket{p_k} = \hat{F}_D\ket{q_k} = \frac{1}{\sqrt{2^N}} \sum_{x_1, \ldots, x_N}
\ket{x_1} \otimes \ket{x_2} \otimes \cdots \otimes \ket{x_N} \, e^{2 \pi i y x/2^N}
\end{equation}
where $y = y_1 \ldots y_N\cdot 1=k+1/2$ and $x = x_1 \ldots x_N\cdot 1=j+1/2$.

The connection with the classical baker's map comes from its symbolic dynamics.
In the quantum case, a string is created through the partial
Fourier transform $\hat{G}_n$.  It is an operator which Fourier transforms the $N-n$ least
significant qubits of a state
\begin{equation}
\label{partialfourier}
\hat{G}_n \Big(\ket{x_1} \otimes \cdots \otimes \ket{x_n} \otimes \ket{a_1} \otimes \cdots
\otimes \ket{a_{N-n}}\Big)
\,\equiv\,\ket{x_1} \otimes \cdots \otimes \ket{x_n} \otimes \frac{1}{\sqrt{2^{N-n}}}
\sum_{x_{n+1},\ldots, x_N} \ket{x_{n+1}} \otimes \cdots \otimes\ket{x_N} e^{2 \pi i a x/2^{N-n}}
\end{equation}
where $a$ and $x$ are defined through the binary expansions $a=a_1 \ldots a_{N-n}\cdot 1$
and $x= x_{n+1} \ldots x_N\cdot 1$. In the limiting cases, we have $\hat{G}_0=\hat{F}_{D}$ and 
$\hat{G}_N=i\hat{1}$.

The analogy to the classical case is made clear through the definition
\begin{equation}
\ket{a_{N-n} \ldots a_1 \bullet x_1 \ldots x_n} \equiv \hat{G}_n \Big( \ket{x_1} \otimes \cdots \otimes \ket{x_n} \otimes \ket{a_1} \otimes \cdots
\otimes \ket{a_{N-n}}\Big). \label{partialfourierstates}
\end{equation}
These states form an orthonormal basis and are localized in both position and momentum. 
They are strictly localized
in a position region of width $1/2^n$ centered at
$0\cdot x_1 \ldots x_n 1$, and are roughly localized
in a momentum region of width $1/2^{N-n}$ centered at
$0\cdot a_1 \ldots a_{N-n} 1$.
\begin{figure}[t]
\includegraphics[scale=1]{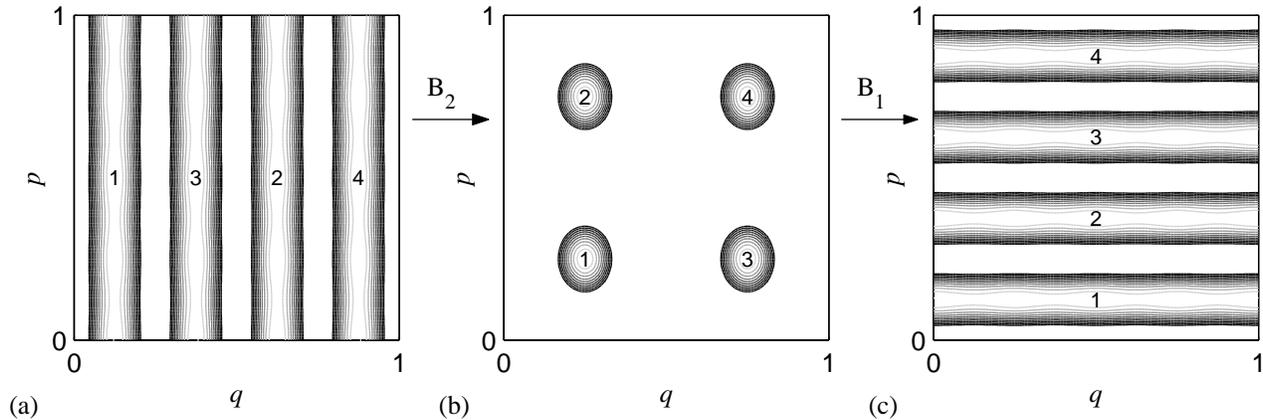}
\caption{The Husimi function for each partially Fourier transformed state 
(\ref{partialfourierstates}) when $N=2$, and, (a) $n=2$, (b) $n=1$ and (c) $n=0$.}
\label{Gstates}
\end{figure}

Using this notation, Schack and Caves defined a whole class of quantum baker's maps 
$\hat{B}_n$ ($n=1, \ldots, N$)
\begin{equation}
\label{bakern}
\hat{B}_n \,\equiv\, \hat{G}^{\phantom{-1}}_{n-1} \circ \hat{G}^{-1}_n \; = \sum_{\begin{array}{c} \scriptstyle x_1,\dots,x_n \\
\scriptstyle a_1,\dots,a_{N-n} \end{array}}
\ket{a_{N-n}\dots a_1 x_1\bullet x_2\dots x_n} \bra{a_{N-n}\dots a_1
\bullet x_1 x_2\dots x_n} .
\end{equation}
The Balazs-Voros-Saraceno quantum baker's map is recovered when $n=1$. In the language of 
Eq. (\ref{symbseq}), we see that each quantum baker's map takes a state localized at 
$1 a_{N-n} \ldots a_1 \bullet x_1 \ldots x_n 1$ to a state localized at 
$1 a_{N-n} \ldots a_1 x_1 \bullet x_2 \ldots x_n 1$.  The decrease in the
number of position bits and increase in momentum bits enforces a stretching and squeezing of 
phase space in a manner resembling the classical baker's map. In figures \ref{Gstates}(a), (b) and 
(c), we have plotted the Husimi function (defined below) for the partially Fourier transformed states 
(\ref{partialfourierstates}) when $N=2$, and $n=2$, 1 and 0, respectively. The quantum baker's map 
is simply a one-to-one mapping of one basis to another. 

It will be useful to rewrite our baker's map in the position basis.  To do this, we
first use (\ref{partialfourier}) and (\ref{partialfourierstates}) to rewrite Eq. (\ref{bakern}) as
\begin{eqnarray}
\hat{B}_n &=& \frac{\sqrt{2}}{2^{N-n+1}}\sum_{\begin{array}{c} \scriptstyle x_1,\dots,x_n \\
\scriptstyle a_1,\dots,a_{N-n} \end{array}}\sum_{\begin{array}{c}
\scriptstyle z_1,\dots,z_{N-n+1} \\ \scriptstyle y_1,\dots,y_{N-n} \end{array}}
\ket{\bullet x_2 \dots x_n z_1 \dots z_{N-n+1}}
\bra{\,\bullet\, x_1 \dots x_n y_1 \dots y_{N-n}} \\ \nonumber
&\phantom{=}&\qquad \times \exp\left[\frac{\pi i}{2^{N-n}}\Big((j+1/2)(l+1/2)+2^{N-n}x_1(l+1/2)
-2(j+1/2)(k+1/2)\Big)\right]
\end{eqnarray}
where 
\[
j = \sum_{k=1}^{N-n}a_k 2^{N-n-k} \text{,}\quad k = \sum_{k=1}^{N-n}y_k 2^{N-n-k}
\quad\text{and}\quad l = \sum_{k=1}^{N-n+1}z_k 2^{N-n+1-k}.
\]
Next, using (\ref{qj}), (\ref{partialfourierstates}) and the notation $q_j = (j+1/2)/D$,
$q_k = (k+1/2)/D$, {\it etc}, we arrive at the quantum baker's map in the position basis
\begin{eqnarray}
\label{bake}
\hat{B}_n &=& \frac{\sqrt{2}}{2^{N-n+1}}\:\:\sum_{x_1=0}^1\:\:\:\sum_{j,k=0}^{2^{N-n}-1}\:\:\:
\sum_{l=0}^{2^{N-n+1}-1}\:\:\:\sum_{m=0}^{2^{n-1}-1} \nonumber\\
&\phantom{=}&\qquad\times\quad \Big|q_l+q_m2^{N-n+1}-2^{-n}\Big\rangle
\Big\langle q_k+x_1/2+q_m2^{N-n}-2^{-n-1}\Big| \nonumber\\
&\phantom{=}& \qquad\qquad\times\quad
\exp\left[i\pi D 2^n\left(q_jq_l+2^{-n}x_1q_l-2q_jq_k\right)\right].
\end{eqnarray}
Note that it is possible to sum over the index $j$ at this point. However the above representation 
proves to be most convenient when performing our semi-classical analysis.   

We will now introduce {\it coherent states} for ${\cal H}_D$ 
\cite{Jin1985a,Leboeuf1990a,Nonnenmacher1998a}:
\begin{eqnarray}
\label{cohnorm}
\ket{a} &\equiv& \frac{1}{\cal N}\left(\frac{2}{D}\right)^{1/4} \msum{\mu} \Dsum{j}
\exp\left[-\frac{\pi D}{2}\left(|a|^2-a^2\right)-\pi D\left(q_j-a+\mu\right)^2+i\pi\mu\right]
\ket{q_j} \\
&=& \frac{1}{\cal N}\left(\frac{2}{D}\right)^{1/4} \;\Dsum{j}
\exp\left[-\frac{\pi D}{2}\left(|a|^2+a^2\right)-\pi D\left(q_j^2-2q_ja\right)\right]
\theta_0\big[iD(q_j-a)\big|iD\big]\ket{q_j}
\end{eqnarray}
where $a\equiv q+ip$ and $\theta_0$ is called a {\it theta function} \cite{akhiezer}
\begin{equation}
\theta_0\big[z\big|\tau\big]\equiv\msum{\mu}\exp\big[i\pi\big(\tau\mu^2+(2z+1)\mu\big)\big].
\end{equation}
The coherent states obey $|a\pm 1\rangle = -\exp[\pm\pi iDp ] |a\rangle $, 
$|a\pm i\,\rangle = -\exp[\mp\pi iDq ] |a\rangle$, and are simply the 
standard (Weyl group) coherent states that have been 
(anti-) periodicized and then projected onto ${\cal H}_D$. The 
normalization factor takes the form 
\begin{equation}
{\cal N}^2 = \theta_0\big[qD\big|iD/2\big]\theta_0\big[pD\big|iD/2\big] = 1+O(1/D) \qquad (D \text{ even})
\end{equation}
and henceforth, will be set to unity. Finally, the {\it Husimi function} for our toroidal 
phase space is defined as $|\langle\psi|a\rangle|^2$.

\section{The Semi-Classical Propagator}
\label{propsect}

\begin{figure}[t]
\includegraphics[scale=0.85]{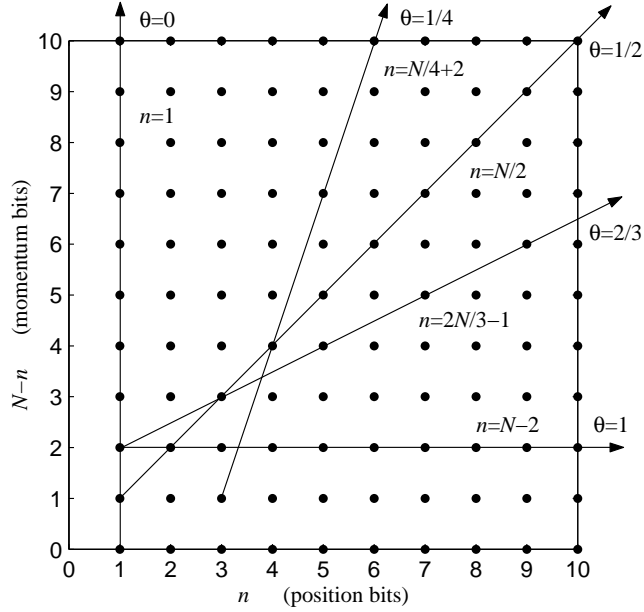}
\caption{Different possible ways of taking the classical limit for the quantum baker's map.}
\label{limits}
\end{figure}

Our goal in this section is to explicitly calculate the semi-classical propagator in the 
coherent state basis. That is, we wish to obtain the leading term in an asymptotic expansion of the matrix 
element $\bra{b}\hat{B}_n\ket{a}$ as $D\rightarrow\infty$.
Observe from Eq. (\ref{bakern}) that in this limit, the total number of position and momentum bits
$N$ necessarily become infinite. However one has considerable freedom of choice 
on how this may occur (see Fig. \ref{limits}). We wish to consider cases where the relative number of position and 
momentum bits approach infinity at different rates. To this end, we take the number of 
position bits to be in the explicit form $n=n(N) \equiv \theta N + s$, 
where $0\leq\theta\leq 1$ is rational and $s$ takes integer values. For ease of reading, we also introduce the 
constant $\phi=1-\theta$ such that the number of momentum bits $N-n = \phi N - s$. 
We will also now identify the different quantum baker's maps through the new parameters, 
$\hat{B}_{\theta,s}\equiv\hat{B}_n$. The parameter $\theta$ ($\phi$) may be interpreted as the 
fraction of qubits allocated to the position (momentum) register as the total number of qubits 
$N$, is increased. In the analysis which follows we must consider the two cases $\theta=0$ and 
$\theta=1$ separately. The former contains the original Balazs-Voros-Saraceno quantization 
($n=s=1$) and will be investigated first. The second parameter $s$, describes an initial offset 
between the number of position and momentum qubits and has no semi-classical effect when $\theta<1$. 
We will find, however, that $s$ becomes important when $\theta=1$.

\subsection{Case $\theta=0$.}

In this case the number of position bits remains constant $n=s\geq 1$ as we let 
$D\rightarrow\infty$. Using (\ref{bake}) and (\ref{cohnorm}) our matrix element becomes
\begin{eqnarray}
\bra{b}\hat{B}_{0,s}\ket{a} &=& SD^{-3/2}\:\msum{\mu,\nu}\:\:\:\sum_{x_1=0}^1\:\:\:\sum_{j,k=0}^{D/S-1}\:\:\:
\sum_{l=0}^{2D/S-1}\:\:\:\sum_{m=0}^{S/2-1}\:\:\: \nonumber\\
&&\exp\bigg[-\frac{\pi D}{2}\Big(|a|^2+|b|^2-a^2-b^{*2}\Big)+i\pi(\mu-\nu)
-\pi D\Big(q_k+x_1/2+(Dq_m-1/2)/S-a+\mu\Big)^2 \nonumber\\
&&-\;\pi D\Big(q_l+2(Dq_m-1/2)/S-b^*+\nu\Big)^2
+i\pi SD\Big(q_jq_l+x_1q_l/S-2q_jq_k\Big)\bigg]
\end{eqnarray}
where $S\equiv 2^s$. To further the calculation, we now use variants of the Poisson summation 
formula to replace each sum with $D$ in the upper limit, by an 
integral {\it e.g.} 
\begin{equation}
\msum{\alpha} \int_0^{1/S} \exp\Big[2\pi i (D x - 1/2)\alpha \Big] f(x)dx
= \frac{1}{D} \int_0^{1/S} \msum{j} \delta\Big(x-(j+1/2)/D\Big) f(x)dx
= \frac{1}{D} \sum_{j=0}^{D/S-1} f(q_j) . \label{sum2int}
\end{equation}
The result is
\begin{eqnarray}
\label{matrixelement}
\bra{b}\hat{B}_{0,s}\ket{a} &=& SD^{3/2}\mathop{\sum_{\mu,\nu,\alpha}^\infty}_{\beta,\gamma=-\infty}
\:\:\:\sum_{x_1=0}^1\:\:\:\sum_{m=0}^{S/2-1}\:\:\: 
\int_0^{1/S} dx \int_0^{1/S} dy \int_0^{2/S} dz \nonumber\\
&&\exp\bigg[-\frac{\pi D}{2}\Big(|a|^2+|b|^2-a^2-b^{*2}\Big)+i\pi(\mu-\nu-\alpha-\beta-\gamma) \nonumber\\
&&-\;\pi D\Big(y+x_1/2+m/S-a+\mu\Big)^2 -\pi D\Big(z+2m/S-b^*+\nu\Big)^2 \nonumber\\
&&+\;i\pi SD\Big(xz+x_1z/S-2xy\Big)+2i\pi D\left(x\alpha+y\beta+z\gamma\right)\bigg].
\end{eqnarray}
We are now ready to make a semi-classical approximation to our matrix element. More 
precisely, we will make a saddle-point approximation to the triple integral above. Only 
near a saddle-point will contributions from such an integral cancel the prefactor 
$D^{3/2}$ and lead to an $O(1)$ contribution for the matrix element. 
The saddle-point approximation can be written down immediately using well-known formulae found 
in any standard text \cite{wong}. However the limits in the above integrals are finite; therefore, 
the saddle point will not make a contribution in all cases. We need to 
consider this possibility carefully if we are to recover the classical baker's map. Hence, we will 
treat each one-dimensional integral separately and use the method of steepest descents. 

Consider first the $y$ integration (with $x$ a parameter) by defining
\begin{equation}
I_1 \equiv \int_0^{1/S} dy\: \exp\big[-\pi D f(y)\big] \label{intA1}
\end{equation}
where
\begin{eqnarray}
f(y) &\equiv& (y-A)^2 + 2i(Sx-\beta)y \\
A &\equiv& a - x_1/2-m/S-\mu.
\end{eqnarray}
\begin{figure}[t]
\includegraphics[scale=0.9]{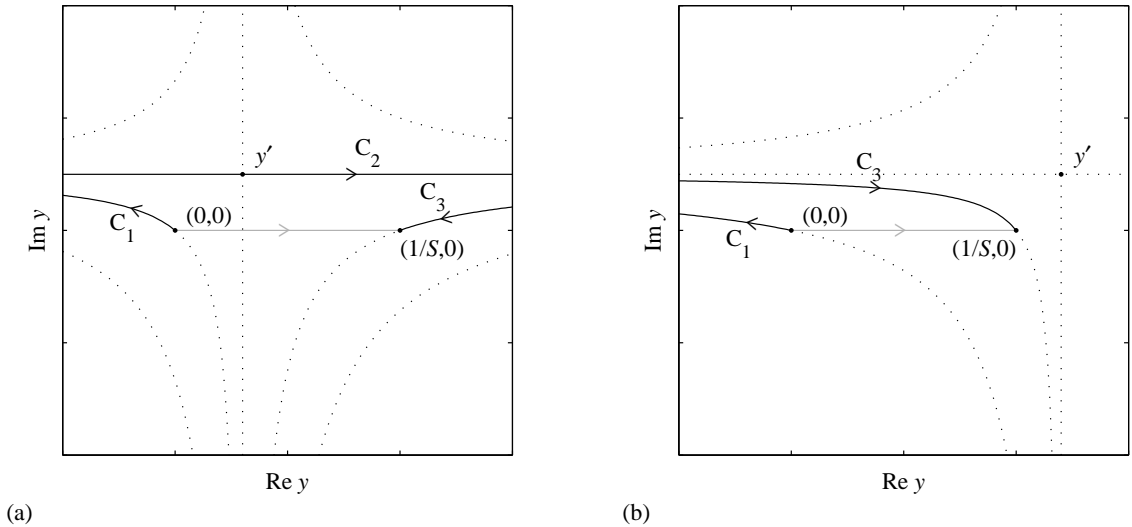}
\caption{Steepest descent paths for $f(y)$. The original integration path along the real line (gray) 
is deformed to one where $\im f(y)=\text{constant}$ (black).}
\label{sdpic}
\end{figure}
An asymptotic calculation of this integral is enabled by deforming the integration path, currently 
along the real line, to one in the complex plane where $\im f(y)=\text{constant}$. Two important 
cases are drawn in Fig. \ref{sdpic}. The first (a) occurs when the saddle point (defined through $f'(y')=0$)
\begin{equation}
y'= A-iSx+i\beta
\end{equation}
satisfies $0<\re y'<1/S$. In this case, the steepest descent path is 
one which first travels along the hyperbola $C_1$ from $0$ to $-\infty+i\im y'$, then along the 
hyperbolic asymptote $C_2$ to $\infty+i\im y'$, and finally back to $1/S$ {\it via} another 
hyperbola $C_3$. Hence an asymptotic expansion for the integral $I_1$ will be the sum of three parts, each 
associated with a different contour $C_1$, $C_2$ or $C_3$ in the complex plane. Note that along the contours $C_1$ and 
$C_3$, the kernel attains its maximum at the end points $0$ and $1/S$, respectively. Consequently, 
the leading term in an asymptotic expansion takes the form
\begin{equation} 
\frac{-1}{\pi D f'(c)}\exp\big[-\pi D f(c)\big]\big[1+O(1/D)\big]
\end{equation}
with $c=0$ or $1/S$. However the prefactor of $1/D$ in the above inhibits such terms from 
playing a role in the leading order approximation of our matrix element. 
As remarked before, we need prefactors of $D^{-1/2}$ in each approximation of the three 
integrals in (\ref{matrixelement}) in order to obtain an $O(1)$ overall contribution for the 
matrix element. Hence we will simply discard the integration along contours $C_1$ and 
$C_3$, and make the approximation
\begin{eqnarray}
I_1 &\approx& \int_{-\infty+i\im y'}^{\infty+i\im y'} dy\: \exp\big[-\pi D f(y)\big] \qquad\text{if}\qquad 0<\re y'<1/S\\
 &=& D^{-1/2} \exp\big[-\pi D f(y')\big] \label{saddle}.
\end{eqnarray} 

When $\re y'<0$ or $\re y'>1/S$ (Fig. \ref{sdpic}(b)) the path of 
steepest descent no longer passes through the saddle point, and consequently, there will be no 
leading order contributions to the matrix element, {\it i.e.} we may set $I_1=0$. The 
third and final case occurs when $\re y'=0$ or $1/S$. One may investigate 
these possibilities by taking exactly one half of (\ref{saddle}) as the approximation for $I_1$. However, 
for simplicity, we will not deal with this case, except make the odd casual remark when 
needed. In summary, we take (\ref{saddle}) as our approximation for $I_1$ when 
$0<\re y'<1/S$, and otherwise zero.

Similarly, for the $z$ integration one has
\begin{eqnarray}
I_2 &\equiv& \int_0^{2/S} dz\: \exp\big[-\pi D g(z)\big] \\
&\approx& D^{-1/2} \exp\big[-\pi D g(z')\big] \qquad\text{if}\qquad 0<\re z'<2/S
\end{eqnarray}
with
\begin{eqnarray}
g(z) &\equiv& (z-B)^2 - 2i(Sx/2+x_1/2+\gamma)z \\
B &\equiv& b^* -2m/S-\nu
\end{eqnarray}
and the saddle point is 
\begin{equation}
z'= B+iSx/2+ix_1/2+i\gamma.
\end{equation}
Now letting $x$ vary again and setting 
\begin{eqnarray}
h(x)&\equiv& f(y')+g(z')-2i\alpha x \\
 &=& \frac{5}{4}S^2x^2+S\Big(2iA-iB+x_1/2-2\beta+\gamma-2i\alpha/S\Big)x 
 -2iA\beta-iB(x_1+2\gamma)+\beta^2+(x_1/2+\gamma)^2
\end{eqnarray}
we have the final integral
\begin{eqnarray}
I_3 &\equiv& \int_0^{1/S} dx\: \exp\big[-\pi D h(x)\big] \\
&\approx& \sqrt{\frac{4}{5S^2D}} \exp\big[-\pi D h(x')\big] \qquad\text{if}\qquad 0<\re x'<1/S
\end{eqnarray}
with the saddle point at
\begin{equation}
x'= -\frac{2}{5S}\Big(2iA-iB+x_1/2-2\beta+\gamma-2i\alpha/S\Big).
\end{equation}

Now, inserting our saddle-point approximations back into (\ref{matrixelement}) and setting
$a\equiv a_1+ia_2$ and $b\equiv b_1+ib_2$, with a little algebra we obtain
\begin{eqnarray}
\label{matrixelement2}
\bra{b}\hat{B}_{0,s}\ket{a} &\approx& \sqrt{\frac{4}{5}}\mathop{\sum_{\mu,\nu,\alpha}^\infty}_{\beta,\gamma=-\infty}
\:\sum_{x_1=0}^1\:\:\sum_{m=0}^{S/2-1}\:\: \exp\bigg[-\frac{\pi D}{5}\Big\{(2a_1-b_1-x_1-2\mu+\nu-2\alpha/S)^2 \nonumber\\
&& +\;(a_2-2b_2+x_1+\beta+2\gamma)^2\Big\} +i\pi(\mu-\nu-\alpha-\beta-\gamma)+i\pi D(a_1a_2-b_1b_2) \nonumber\\
&&-\;\frac{2i\pi D}{5}\Big\{(2a_1-b_1-x_1-2\mu+\nu-2\alpha/S)(2a_2+b_2) \nonumber\\
&&-\;(a_1+2b_1-x_1/2-\mu-2\nu)(x_1+\beta+2\gamma)+\alpha(x_1-4\beta+2\gamma)/S\Big\}\bigg]
\end{eqnarray} 
provided that all three of the inequalities
\begin{eqnarray}
0&<&\re x' =\frac{2}{5S}\Big(2a_2+b_2-x_1/2+2\beta-\gamma\Big)<\frac{1}{S} \label{ineq1}\\
0&<&\re y' =\frac{1}{5}\Big(a_1+2b_1-x_1/2-\mu-2\nu\Big)+\frac{1}{5S}\Big(4\alpha-5m\Big)<\frac{1}{S} \label{ineq2}\\
0&<&\frac{1}{2}\re z' = \frac{1}{5}\Big(a_1+2b_1-x_1/2-\mu-2\nu\Big)-\frac{1}{5S}\Big(\alpha+5m\Big)<\frac{1}{S} \label{ineq3}
\end{eqnarray}
are satisfied. Otherwise the summand is taken to be zero. 

Note that although $m$ no longer appears in the exponent, we cannot trivially evaluate the sum 
since not all values of 
$m$ will satisfy (\ref{ineq2}) and (\ref{ineq3}). Consider the cases when the approximation
(\ref{matrixelement2}) becomes $O(1)$. That is,
\begin{eqnarray}
2a_1-b_1-x_1-2\mu+\nu-2\alpha/S &=& 0 \label{O1approx1}\\
a_2-2b_2+x_1+\beta+2\gamma &=& 0. \label{O1approx2}
\end{eqnarray}
Substituting (\ref{O1approx2}) into (\ref{ineq1}) we obtain 
\begin{equation}
0<a_2+\beta<1 \qquad\text{or}\qquad 0<b_2-x_1/2-\gamma<1/2 \label{newineq1}
\end{equation}
and thus, the integers $\beta$ and $\gamma$ give our periodicity in the momentum direction. Hence 
if we assume $0<a_2,b_2<1$ then we may set $\beta=\gamma=0$ in (\ref{matrixelement2}), making 
note that we are discarding exponentially small Gaussian tails. Also note from (\ref{newineq1}) 
that we must have $x_1=\lfloor 2b_2 \rfloor$. 

Now, negating (\ref{ineq3}) and adding it to (\ref{ineq2}) we immediately arrive at the 
inequality $-1<\alpha<1$. Hence we must set $\alpha=0$. This implies 
\begin{equation}
\frac{2m}{S}<\frac{2}{5}\Big(a_1+2b_1-x_1/2-\mu-2\nu\Big)<\frac{2(m+1)}{S} \label{newineq2}
\end{equation}
from (\ref{ineq2},\ref{ineq3}), or equivalently
\begin{equation}
0<\frac{2}{5}\Big(a_1+2b_1-x_1/2-\mu-2\nu\Big)<1 \label{newineq3}
\end{equation}
if we now drop the summation over $m$ in (\ref{matrixelement2}). Hence, following a similar procedure to the 
above, one can substitute (\ref{O1approx1}) into the new inequality (\ref{newineq3}) and 
deduce that under the assumption $0<a_1,b_1<1$, the summand of (\ref{matrixelement2}) becomes 
$O(1)$ only when $\mu=\nu=0$. Furthermore, we will have $x_1=\lfloor 2a_1 \rfloor$.

The surviving term of the summation is our semi-classical 
approximation for the propagator: 
\begin{eqnarray}
\bra{b}\hat{B}_{0,s}\ket{a} &=& \sqrt{\frac{4}{5}} \exp\bigg[-\frac{\pi D}{5}\Big\{\big(2a_1-b_1-\lfloor 2a_1\rfloor\big)^2 +\big(a_2-2b_2+\lfloor 2a_1\rfloor\big)^2 \nonumber\\
&&+\;i\big(3a_1a_2+3b_1b_2+4a_1b_2-4a_2b_1\big)-2i\lfloor 2a_1\rfloor\big(a_1+2b_1+2a_2+b_2-\lfloor 2a_1\rfloor/2\big)\Big\}\bigg]+o(1) \label{semiS1}
\end{eqnarray} 
where we have chosen $x_1=\lfloor 2a_1 \rfloor$ (and implicitly assumed $a_1\neq 1/2$ and 
$b_2\neq 1/2$ by ignoring cases of equality in (\ref{ineq1}-\ref{ineq3})). All other terms 
in (\ref{matrixelement2}), being exponentially small, are discarded. 

Note that the above 
approximation is $O(1)$ only when $b$ is the iterate of $a$ under the classical baker's map 
(\ref{classbake3}). Furthermore, a little algebra reveals that our semi-classical propagator 
may be rewritten in the Van Vleck form
\begin{equation}
\bra{b}\hat{B}_{0,s}\ket{a} = \sqrt{\frac{\partial^2 W}{\partial a\partial b^*}}
\exp\Big[\pi D W(b^*,a)\Big]\exp\Big[-\pi D\left(|a|^2+|b|^2\right)/2\Big] +o(1) \label{vanvleck}
\end{equation}
where $W(b^*,a)$ is the classical generating function (\ref{genfunc}). Hence we have shown that 
the class of quantum baker's map with $\theta=0$ will approach the classical baker's map in the 
limit $D\rightarrow\infty$.

\subsection{Case $0<\theta<1$.}

We will now consider the case $0<\theta<1$. Using (\ref{bake}) and (\ref{cohnorm}) with 
$n=\theta N + s$, our matrix element is
\begin{eqnarray}
\bra{b}\hat{B}_{\theta,s}\ket{a} &=& \frac{S}{\sqrt{D}D^\phi}\:\msum{\mu,\nu}\:\:\:\sum_{x_1=0}^1\:\:\:\sum_{j,k=0}^{D^\phi/S-1}\:\:\:
\sum_{l=0}^{2D^\phi/S-1}\:\:\:\sum_{m=0}^{SD^\theta/2-1}\:\:\: \nonumber\\
&&\exp\bigg[-\frac{\pi D}{2}\Big(|a|^2+|b|^2-a^2-b^{*2}\Big)+i\pi(\mu-\nu)
-\pi D\Big(\big(q^\phi_k-1/2S\big)/D^\theta+x_1/2+q^\theta_m/S-a+\mu\Big)^2 \nonumber\\
&&-\;\pi D\Big(\big(q^\phi_l-1/S\big)/D^\theta+2q^\theta_m/S-b^*+\nu\Big)^2
+i\pi SD^\phi\Big(q^\phi_jq^\phi_l+x_1q^\phi_l/S-2q^\phi_jq^\phi_k\Big)\bigg]
\end{eqnarray}
where again $S\equiv 2^s$. Introducing the new summing variables $q^\theta_m\equiv q_mD/D^\theta$,
$q^\phi_j\equiv q_jD/D^\phi$, {\it etc.}, enables us to convert the four finite sums over $j$, $k$, $l$ 
and $m$, to integrals over $x$, $y$, $z$ and $t$, respectively, using formulae similar to 
(\ref{sum2int}). The result is  
\begin{eqnarray}
\bra{b}\hat{B}_{\theta,s}\ket{a} &=& S\sqrt{D}D^\phi\:\mathop{\sum_{\mu,\nu,\alpha,\beta}^\infty}_{\gamma,\kappa=-\infty} \:\:\:\sum_{x_1=0}^1
\:\:\:\int_0^{1/S} dx \int_0^{1/S} dy \int_0^{2/S} dz \int_0^{S/2} dt\nonumber\\
&&\exp\bigg[-\frac{\pi D}{2}\Big(|a|^2+|b|^2-a^2-b^{*2}\Big)+i\pi(\mu-\nu-\alpha-\beta-\gamma-\kappa) \nonumber\\
&&-\;\pi D\Big(\big(y-1/2S\big)/D^\theta+x_1/2+t/S-a+\mu\Big)^2 -\pi D\Big(\big(z-1/S\big)/D^\theta+2t/S-b^*+\nu\Big)^2 \nonumber\\
&&+\;i\pi SD^\phi\Big(xz+x_1z/S-2xy\Big)+2i\pi D^\phi(x\alpha+y\beta+z\gamma)+2i\pi D^\theta t\kappa\bigg] \\
&=& \frac{\sqrt{D}D^\phi}{S}\:\mathop{\sum_{\mu,\nu,\alpha,\beta}^\infty}_{\gamma,\kappa=-\infty} \:\:\:\sum_{x_1=0}^1
\:\:\:\int_0^1 dx \int_0^1 dy \int_0^2 dz \int_0^{1/2} dt\nonumber\\
&&\exp\bigg[-\frac{\pi D}{2}\Big(|a|^2+|b|^2-a^2-b^{*2}\Big)+i\pi(\mu-\nu-\alpha-\beta-\gamma-\kappa) \nonumber\\
&&-\;\pi D\big(t+x_1/2-a+\mu\big)^2 -\pi D\big(2t-b^*+\nu\big)^2 \nonumber\\
&&-\;2\pi D^\phi\big(y-1/2\big)\big(t+x_1/2-a+\mu\big)/S -2\pi D^\phi\big(z-1\big)\big(2t-b^*+\nu\big)/S \nonumber\\
&&+\;i\pi D^\phi\big(xz+x_1z-2xy\big)/S+2i\pi D^\phi\big(x\alpha+y\beta+z\gamma\big)/S \nonumber\\
&&-\;\pi D^{\phi-\theta}\big(y-1/2\big)^2/S^2 -\pi D^{\phi-\theta}\big(z-1\big)^2/S^2+2i\pi SD^\theta t\kappa\bigg] \label{monster}
\end{eqnarray}
where we have rescaled the integration variables ($x\rightarrow x/S$, $y\rightarrow y/S$, 
$z\rightarrow z/S$ and $t\rightarrow tS$), then collected terms in the exponent with the same 
power of $D$. The terms with highest power are those containing $t$, and hence, we will 
consider the integration over this variable first. Define the integral 
\begin{equation}
I_4 \equiv \int_0^{1/2} dt\: \exp\Big[-\pi D(t-A)^2 -\pi D(2t-B)^2
-2\pi D^\phi t C+ 2i\pi SD^\theta t \kappa \Big] \label{int4}
\end{equation}
where the constants are
\begin{eqnarray}
\label{ABCEdef}
A &\equiv& a-x_1/2-\mu \\ 
B &\equiv& b^*-\nu \\ 
C &\equiv& (y-1/2)/S+2(z-1)/S.
\end{eqnarray}
We now wish to derive the contribution from $I_4$ which gives the leading order approximation 
to our matrix element. This is done by taking a path of steepest descent for the function 
$f(t)\equiv (t-A)^2+(2t-B)^2$. Note that when $\phi=1$ the third term of the exponent in (\ref{int4}) 
also becomes dominant and must be incorporated into $f(t)$. Hence the need to consider this case 
separately in the previous section. 

As before, we may discard all parts of our integration contour, except the segment $(-\infty+i\im t',\infty+i\im t')$
which passes through the saddle point
\begin{equation}
t'=\frac{A+2B}{5}.
\end{equation}
It is only this contribution which will cancel the prefactor $\sqrt{D}D^\phi$ in 
(\ref{monster}) to give an $O(1)$ overall contribution to the matrix element. Hence we make 
the approximation  
\begin{eqnarray}
I_4 &\approx& \int_{-\infty+i\im t'}^{\infty+i\im t'} dt\: \exp\Big[-\pi D(t-A)^2 -\pi D(2t-B)^2
-2\pi D^\phi t C+ 2i\pi SD^\theta t \kappa \Big] \\
&=& \frac{1}{\sqrt{5D}} \exp\bigg[-\frac{\pi D}{5}(2A-B)^2-\frac{2\pi D^\phi}{5}(A+2B)C+\frac{2i\pi D^\theta}{5}(A+2B)S\kappa \nonumber\\
&&+\;\frac{\pi D^{\phi-\theta}}{5}C^2-\frac{\pi D^{\theta-\phi}}{5}S^2\kappa^2-\frac{2i\pi}{5}CS\kappa\bigg] \label{int4approx}
\end{eqnarray} 
if $0<\re t'<1/2$, and otherwise zero. 

Substituting this approximation back into (\ref{monster}) and simplifying we obtain
\begin{eqnarray}
\bra{b}\hat{B}_{\theta,s}\ket{a} &\approx& \frac{D^\phi}{\sqrt{5}S}\:\mathop{\sum_{\mu,\nu,\alpha,\beta}^\infty}_{\gamma,\kappa=-\infty} \:\:\:\sum_{x_1=0}^1
\:\:\:\int_0^1 dx \int_0^1 dy \int_0^2 dz \nonumber\\
&&\exp\bigg[-\frac{\pi D}{2}\Big(|a|^2+|b|^2-a^2-b^{*2}\Big)+i\pi(\mu-\nu-\alpha-\beta-\gamma)- \frac{\pi D}{5}\big(2A-B\big)^2 \nonumber\\
&&+\;\frac{2\pi D^\phi}{5S}\big(2A-B\big)\big(2y-z\big)+\frac{i\pi D^\phi}{S}\big(xz+x_1z-2xy\big)+\frac{2i\pi D^\phi}{S}\big(x\alpha+y\beta+z\gamma\big) \nonumber\\
&&-\;\frac{\pi D^{\phi-\theta}}{5S^2}\big(2y-z\big)^2-\frac{2i\pi}{5}\big(y+2z\big)\kappa+\frac{2i\pi D^\theta}{5}\big(A+2B\big)S\kappa-\frac{\pi D^{\theta-\phi}}{5}S^2\kappa^2\bigg]. \label{elmess}
\end{eqnarray}
The dominant terms in the exponent which contain the integration variables, are now those with $D^\phi$ 
as a prefactor. These terms do not define a saddle point, but instead, a line. Hence, it is 
advantageous to first decouple $x$, $y$ and $z$ in these terms using the following transformation
\begin{equation}
\label{variablechange}
\left[ \begin{array}{l} u \\ v \\ w \end{array} \right] =
\left[ \begin{array}{rrr} 1 & -2 & 1 \\ 1 & 2 & -1 \\ 0 & 1 & 2 \end{array} \right]
\left[ \begin{array}{l} x \\ y \\ z \end{array} \right] 
\end{equation}
where the integration region $(x,y,z) \in [0,1]\times [0,1]\times [0,2]$ is transformed to some 
parallel-piped $\Omega$.

After making this transformation, Eq. (\ref{elmess}) may be rewritten in the form
\begin{equation}
\bra{b}\hat{B}_{\theta,s}\ket{a} \approx \frac{D^\phi}{10 \sqrt{5}S}
\mathop{\sum_{\mu,\nu,\alpha,\beta}^\infty}_{\gamma,\kappa=-\infty}\:\: \sum_{x_1=0}^1 \:\:
\mathop{\int\!\!\!\!\int\!\!\!\!\int}_{\Omega} du\, dv\, dw \, \exp\bigg[- \frac{\pi D^\phi}{S} g(u) - \frac{\pi D^\phi}{S} h(v)
-\frac{\pi D^{\phi-\theta}}{20S^2} (u-v)^2\bigg] F(w) \label{monster3}
\end{equation}
where
\begin{eqnarray}
g(u) &\equiv& -\frac{i}{4}u^2-(E+i\alpha)u \\ 
h(v) &\equiv&  \frac{i}{4}v^2+(E-i\alpha)v \label{h}\\ 
E &\equiv& -\frac{1}{5}\big(2A-B\big)+\frac{i}{5}\big( x_1/2-2\beta+\gamma\big)\equiv E_1 + i E_2 \\ \nonumber
\end{eqnarray}
and
\begin{eqnarray}
F(w) &\equiv& \exp\bigg[-\frac{\pi D}{2}\Big(|a|^2+|b|^2-a^2-b^{*2}\Big)+i\pi(\mu-\nu-\alpha-\beta-\gamma)- \frac{\pi D}{5}\big(2A-B\big)^2 \nonumber \\
&&+\;\frac{2i\pi D^\theta}{5}\big(A+2B\big)S\kappa-\frac{\pi D^{\theta-\phi}}{5}S^2\kappa^2+\frac{2i\pi}{5S}\Big(D^\phi(x_1+\beta+2\gamma)-S\kappa\Big)w\bigg].
\end{eqnarray}
\begin{figure}[t]
\includegraphics[scale=0.9]{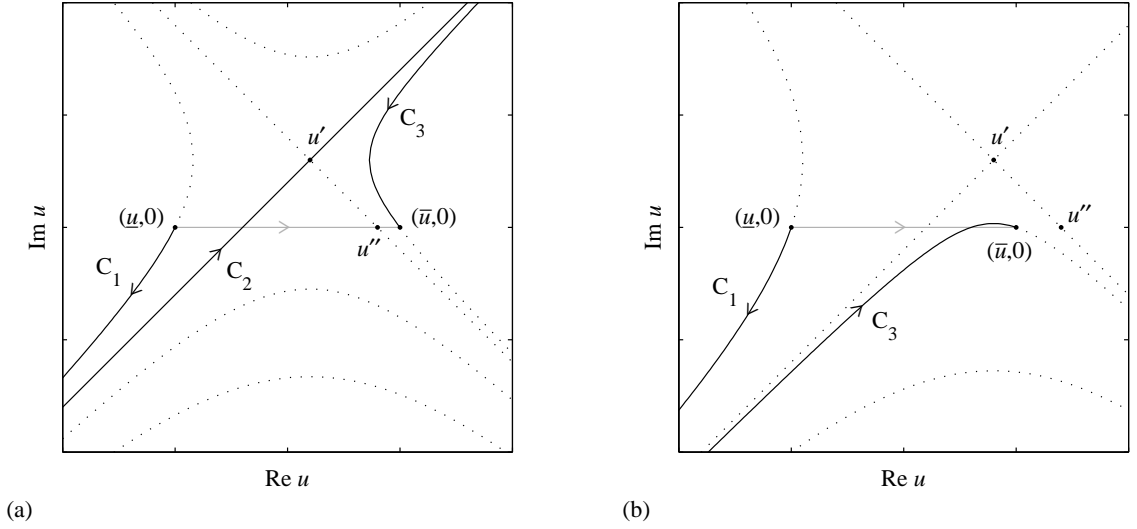}
\caption{Steepest descent paths for $g(u)$.}
\label{sdpic2}
\end{figure}

We have now arrived at a form where we can consider steepest descent paths for the 
integration variables $u$ and $v$. Starting our program with the function $g(u)$ 
and writing $u$ in terms of its real and imaginary parts $u=u_1+iu_2$, ones finds that the 
two hyperbolic asymptotes
\begin{eqnarray}
u_1-u'_1 &=& u_2-u'_2 \label{asymptote1} \\
u_1-u'_1 &=& -u_2+u'_2\label{asymptote2}
\end{eqnarray}
are the steepest descent paths which pass through the saddle point
\begin{equation}
u' = u_1'+iu_2' = 2(iE-\alpha).\label{saddleg}
\end{equation}
However, only the first (\ref{asymptote1}) can be used as an 
integration contour, since $\exp\left[-\pi D^\phi\re g(u)/S\right]\rightarrow\infty$ on the other.
Cases for when this asymptote (denoted by $C_2$) is required to form part of the integration 
contour, and when it is not, are plotted in figures \ref{sdpic2}(a) and (b), respectively. Here the 
integration limits are denoted by the real numbers $\underline{u}$ and $\overline{u}$, and need not 
be known explicitly for the moment. Note that $C_2$ is included in the contour only when the 
intercept of the second asymptote (\ref{asymptote2}) with the real line, denoted by $u''$, is 
between these limits. That is
\begin{equation}
\label{urest}
\underline{u} < u'' = u'_1+u'_2 = 2(E_1-E_2-\alpha)< \overline{u}.
\end{equation}
The importance of this inequality is not clear yet.  We shall return to it after transforming back
to our $x$, $y$ and $z$ variables.

The analysis of $h(v)$ is very similar. In this case one finds that our steepest descent path will 
travel along the asymptote 
\begin{equation}
v_1-v'_1=-v_2+v'_2 \label{asymptotev}
\end{equation}
and hence, through the saddle point
\begin{equation}
\label{saddleh}
v' = v_1'+iv_2' = 2(iE+\alpha)
\end{equation}
only when
\begin{equation}
\label{vrest}
\underline{v} < v'' = v'_1-v'_2 = 2(\alpha-E_1-E_2) < \overline{v}.
\end{equation}

But what are our integration limits $\underline{u}$, $\overline{u}$, $\underline{v}$ and 
$\overline{v}$?  Unfortunately, given the nature of the variable change, their values differ 
as one integrates over the volume element $\Omega$. Therefore, it is convenient to convert back to 
our original variables $x$, $y$ and $z$ which have independent limits. We know that whenever 
the point $(u'',v'',w)$ belongs to our integration region $\Omega$, the two contours (\ref{asymptote1}) and (\ref{asymptotev}) will be 
included in our integration path of steepest descent. Therefore, by inverting our transformation 
(\ref{variablechange}), we obtain
\begin{equation}
\label{xyzdp}
\left.
\begin{array}{lll} u'' &=& 2(E_1-E_2-\alpha) \\ v'' &=& 2(\alpha-E_1-E_2) \\ w &=& w
\end{array} \right.
\Rightarrow
\left.
\begin{array}{lll} x'' &=& -2E_2 \\ y'' &=& \frac{4}{5}(\alpha-E_1) + \frac{1}{5} w \\
z'' &=& -\frac{2}{5}(\alpha-E_1) + \frac{2}{5} w
\end{array} \right.
\end{equation}
and hence, the condition $(u'',v'',w)\in\Omega$ is equivalent to
\begin{eqnarray}
0 < &x''& < 1 \label{ineqx}\\
0 < &y''& < 1 \label{ineqy}\\
0 < &z''& < 2 . \label{ineqz}
\end{eqnarray}
Inequalities (\ref{ineqy}) and (\ref{ineqz}) may be rewritten as
\begin{eqnarray}
\alpha-E_1 < &w& < 5 + \alpha-E_1 \\
-4(\alpha-E_1) < &w& < 5 - 4(\alpha-E_1) 
\end{eqnarray}
which asserts that 
\begin{equation} 
\underline{w} < w < \overline{w}
\end{equation}
where $\underline{w}\equiv\max\{\alpha-E_1,-4(\alpha-E_1)\}$ and 
$\overline{w}\equiv 5+\min\{\alpha-E_1,-4(\alpha-E_1)\}$. The requirement 
that $\underline{w}<\overline{w}$ implies 
\begin{equation}
\label{alphares}
-1 < \alpha-E_1 < 1.
\end{equation}

Having learned all we can from the method of steepest descents about the restrictions placed 
upon our integration parameters, we may proceed with the saddle-point approximation of 
(\ref{monster3}). This is done by discarding all contours which do not make a contribution 
to the leading order approximation 
of our matrix element (e.g. for $u$, we discard $C_1$ and $C_3$ in Fig. \ref{sdpic2}). Hence 
\begin{eqnarray}
\bra{b}\hat{B}_{\theta,s}\ket{a} &\approx& \frac{D^\phi}{10 \sqrt{5}S}
\mathop{\sum_{\mu,\nu,\alpha,\beta}^\infty}_{\gamma,\kappa=-\infty}\:\: \sum_{x_1=0}^1  
\:\:\:\int_{u'-\infty e^{i\pi/4}}^{u'+\infty e^{i\pi/4}} du 
\:\:\:\int_{v'-\infty e^{-i\pi/4}}^{v'+\infty e^{-i\pi/4}}  dv 
\:\:\:\int_{\underline{w}}^{\overline{w}} dw  \nonumber\\
&& \exp\bigg[- \frac{\pi D^\phi}{S} g(u) - \frac{\pi D^\phi}{S} h(v)
-\frac{\pi D^{\phi-\theta}}{20S^2} (u-v)^2\bigg] F(w) \\
&=& \frac{1}{5}\sqrt{\frac{4}{5}}
\mathop{\sum_{\mu,\nu,\alpha,\beta}^\infty}_{\gamma,\kappa=-\infty}\:\: \sum_{x_1=0}^1
\:\:\int_{\underline{w}}^{\overline{w}} dw \:\:\exp\bigg[-\frac{\pi D}{2}\Big(|a|^2+|b|^2-a^2-b^{*2}\Big)\nonumber\\
&&+\;i\pi(\mu-\nu-\alpha-\beta-\gamma)-\frac{\pi D}{5}\big(2A-B\big)^2-\frac{4\pi D^\phi}{S}E\alpha-\frac{4\pi D^{\phi-\theta}}{5S^2}\alpha^2 \nonumber\\
&&+\;\frac{2i\pi D^\theta}{5}\big(A+2B\big)S\kappa-\frac{\pi D^{\theta-\phi}}{5}S^2\kappa^2+\frac{2i\pi}{5S}\Big(D^\phi(x_1+\beta+2\gamma)-S\kappa\Big)w\bigg]. \label{monster4}
\end{eqnarray}
Notice that we have actually made a saddle-point 
approximation about a line parametrized by $w$, over which, we still need to integrate. 
Currently there are no restrictions placed upon $\kappa$. However the 
integral over $w$ will be $O(D^{-\phi})$ unless $\kappa=D^\phi(x_1+\beta+2\gamma)/S$. 
We may set $\kappa$ to this value by noting that we are only shifting the saddle point $t'$ 
by the imaginary amount $i\pi(x_1+\beta+2\gamma)$ (see Eq. (\ref{int4})),
and thus, our approximation of $I_4$ (\ref{int4approx}) is unchanged. 

Hence, putting $\kappa=D^\phi(x_1+\beta+2\gamma)/S$ and integrating over $w$, we arrive at
\begin{eqnarray}
\label{almostdone}
\bra{b}\hat{B}_{\theta,s}\ket{a} &\approx& \frac{1}{5}\sqrt{\frac{4}{5}}
\mathop{\sum_{\mu,\nu,\alpha,\beta}^\infty}_{\gamma=-\infty}\:\: \sum_{x_1=0}^1 \:\big(\overline{w}-\underline{w}\big)
\exp\bigg[-\frac{\pi D}{5}\Big\{(2a_1-b_1-x_1-2\mu+\nu)^2 \nonumber\\
&& +\;(a_2-2b_2+x_1+\beta+2\gamma)^2\Big\} +i\pi(\mu-\nu-\alpha-\beta-\gamma)+i\pi D(a_1a_2-b_1b_2) \nonumber\\
&&-\;\frac{2i\pi D}{5}\Big\{(2a_1-b_1-x_1-2\mu+\nu)(2a_2+b_2) \nonumber\\
&&-\;(a_1+2b_1-x_1/2-\mu-2\nu)(x_1+\beta+2\gamma)\Big\}-\frac{4\pi D^\phi}{S}E\alpha-\frac{4\pi D^{\phi-\theta}}{5S^2}\alpha^2 \bigg]
\end{eqnarray}
provided that all three of the inequalities
\begin{eqnarray}
0 &<& x'' =\frac{2}{5}\Big(2a_2+b_2-x_1/2+2\beta-\gamma\Big)<1 \label{ineqb1}\\
0 &<& 2\re t' =\frac{2}{5}\Big(a_1+2b_1-x_1/2-\mu-2\nu\Big)<1 \label{ineqb2}\\
-1 &<& \alpha+\frac{1}{5}\Big(2a_1-b_1-x_1-2\mu+\nu\Big)<1 \label{ineqb3}
\end{eqnarray}
are satisfied. Otherwise the summand is taken to be zero.

Again, in a similar fashion to the previous section, we note that 
if the above approximation is to become $O(1)$, we must set 
$2a_1-b_1-x_1-2\mu+\nu=0$ and $a_2-2b_2+x_1+\beta+2\gamma=0$. Consequently, under the assumption 
$0<a_1,a_2,b_1,b_2<1$, the above three inequalities will now require us to set 
$\mu=\nu=\alpha=\beta=\gamma=0$ and $x_1=\lfloor 2a_1 \rfloor$. Furthermore, we are 
assuming $a_1\neq 1/2$ and $b_2\neq 1/2$ by ignoring cases of equality in (\ref{ineqb1}-\ref{ineqb3}). 

Thus, by discarding all exponentially small terms, we arrive at the semi-classical propagator 
\begin{eqnarray}
\bra{b}\hat{B}_{\theta,s}\ket{a} &=& \sqrt{\frac{4}{5}}\bigg(1-\frac{1}{5}\big|2a_1-b_1-\lfloor 2a_1\rfloor\big|\bigg)
\exp\bigg[-\frac{\pi D}{5}\Big\{\big(2a_1-b_1-\lfloor 2a_1\rfloor\big)^2+\big(a_2-2b_2+\lfloor 2a_1\rfloor\big)^2 \nonumber\\
&&+\;i\big(3a_1a_2+3b_1b_2+4a_1b_2-4a_2b_1\big)-2i\lfloor 2a_1\rfloor\big(a_1+2b_1+2a_2+b_2-\lfloor 2a_1\rfloor/2\big)\Big\}\bigg]+o(1) \label{form1}\\
&=& \sqrt{\frac{4}{5}}\exp\bigg[-\frac{\pi D}{5}\Big\{\big(2a_1-b_1-\lfloor 2a_1\rfloor\big)^2+\big(a_2-2b_2+\lfloor 2a_1\rfloor\big)^2 \nonumber\\
&&+\;i\big(3a_1a_2+3b_1b_2+4a_1b_2-4a_2b_1\big)-2i\lfloor 2a_1\rfloor\big(a_1+2b_1+2a_2+b_2-\lfloor 2a_1\rfloor/2\big)\Big\}\bigg]+o(1). \label{form2}
\end{eqnarray}
Both forms, (\ref{form1}) and (\ref{form2}), are equally valid since their difference is exponentially 
small (although the first (\ref{form1}) may prove to be more accurate). Comparing (\ref{form2}) to 
(\ref{semiS1}), we see that for $0<\theta<1$, the semi-classical 
propagator also takes the Van Vleck form (\ref{vanvleck}). 
Furthermore, the classical baker's map will be recovered in the limit $D\rightarrow\infty$.

\subsection{Case $\theta=1$.}

After the rousing success of the previous calculations, it is tempting to conclude 
that the classical baker's map will always be restored in the limit $D\rightarrow\infty$. 
Unfortunately, when $\theta=1$ ($\phi=0$), certain assumptions made previously will prove incorrect. 
In particular, in Eq. (\ref{elmess}), we have assumed that $D^\phi$ terms will dominate; 
however this clearly cannot now be the case. In this section we show that such differences 
cripple any hope that the classical baker's map will be recovered for all possible classical limits.

When $\theta=1$ the number of momentum bits $r\equiv-s\geq 0$ remains constant.
Using (\ref{bake}) and (\ref{cohnorm}), with $n=N-r$ and $R\equiv 2^r$, our matrix element is
\begin{eqnarray}
\bra{b}\hat{B}_{1,-r}\ket{a} &=& \frac{1}{\sqrt{D}R}\:\msum{\mu,\nu}\:\:\:\sum_{x_1=0}^1\:\:\:\sum_{j,k=0}^{R-1}\:\:\:
\sum_{l=0}^{2R-1}\:\:\:\sum_{m=0}^{D/(2R)-1}\:\:\exp\bigg[-\frac{\pi D}{2}\Big(|a|^2+|b|^2-a^2-b^{*2}\Big)+i\pi(\mu-\nu) \nonumber\\
&&-\;\pi D\Big(Rq_m+x_1/2-a+\mu+\big(k+1/2-R/2\big)/D\Big)^2-\pi D\Big(2Rq_m-b^*+\nu+\big(l+1/2-R\big)/D\Big)^2 \nonumber \\
&&+\;\frac{i\pi}{R}\Big\{\big(j+1/2\big)\big(l+1/2\big)+Rx_1\big(l+1/2\big)-2\big(j+1/2\big)\big(k+1/2\big)\Big\}\bigg] \\
&=& \frac{\sqrt{D}}{R^2}\msum{\mu,\nu,\kappa}\:\:\:\sum_{x_1=0}^1\:\:\:\sum_{j,k=0}^{R-1}\:\:\:
\sum_{l=0}^{2R-1}\:\:\int_0^{1/2} dt\:\:\exp\bigg[-\frac{\pi D}{2}\Big(|a|^2+|b|^2-a^2-b^{*2}\Big)+i\pi(\mu-\nu-\kappa) \nonumber\\
&&-\;\pi D\big(t+x_1/2-a+\mu\big)^2-\pi D\big(2t-b^*+\nu\big)^2-\frac{\pi}{D}\big(k+1/2-R/2\big)^2-\frac{\pi}{D}\big(l+1/2-R\big)^2  \nonumber \\
&&+\;\frac{2i\pi D}{R}t\kappa-2\pi\big(t+x_1/2-a+\mu\big)\big(k+1/2-R/2\big)-2\pi\big(2t-b^*+\nu\big)\big(l+1/2-R\big)\nonumber\\
&&+\;\frac{i\pi}{R}\Big\{\big(j+1/2\big)\big(l+1/2\big)+Rx_1\big(l+1/2\big)-2\big(j+1/2\big)\big(k+1/2\big)\Big\}\bigg] \label{matrixel1}
\end{eqnarray}
where we have converted the sum over $m$ to an integral over $t$ using the same technique in the previous sections.

Now defining the integral
\begin{equation}
I_5\equiv\int_0^{1/2}dt\;\exp\Big[-\pi Df(t)-2\pi\big(t-A\big)\big(k+1/2-R/2\big)-2\pi\big(2t-B\big)\big(l+1/2-R\big)\Big]
\end{equation}
where
\begin{eqnarray}
f(t) &\equiv& \big(t-A\big)^2+\big(2t-B\big)^2-2it\kappa/R \\
A &\equiv& a-x_1/2-\mu \\ 
B &\equiv& b^*-\nu
\end{eqnarray}
one finds the saddle point
\begin{equation}
t'=\frac{A+2B}{5}+\frac{i\kappa}{5R}
\end{equation}
and hence, the approximation
\begin{eqnarray}
I_5 &\approx& \int_{-\infty+i\im t'}^{\infty+i\im t'} dt\;\exp\Big[-\pi Df(t)-2\pi\big(t-A\big)\big(k+1/2-R/2\big)-2\pi\big(2t-B\big)\big(l+1/2-R\big)\Big] \\
&=& \frac{1}{\sqrt{5D}}\;\exp\bigg[-\frac{\pi D}{5}\big(2A-B\big)^2+\frac{2i\pi D}{5R}\big(A+2B\big)\kappa-\frac{\pi D}{5R^2}\kappa^2+i\pi\kappa \nonumber\\
&&+\;\frac{2\pi}{5}\big(2A-B\big)\big(2k-l+1/2\big)-\frac{2i\pi}{5R}\big(k+2l+3/2\big)\kappa+\frac{\pi}{5D}\big(k+2l+3/2-5R/2\big)^2 \bigg] \label{I5approx}
\end{eqnarray}
provided that $0<\re t'<1/2$, and otherwise zero. 

Apart from the last term in the exponent, Eq. (\ref{I5approx}) is simply the well-known formula for a saddle-point 
approximation found in any standard text. We will now drop this $1/D$ term, along with all others 
in (\ref{matrixel1}), to obtain the approximation 
\begin{eqnarray}
\bra{b}\hat{B}_{1,-r}\ket{a} &\approx& \frac{1}{\sqrt{5}R^2}\msum{\mu,\nu,\kappa}\:\:\:\sum_{x_1=0}^1\:\:\:\sum_{j,k=0}^{R-1}\:\:\:
\sum_{l=0}^{2R-1}\:\:\exp\bigg[-\frac{\pi D}{5}\Big\{\big(2a_1-b_1-x_1-2\mu+\nu\big)^2 \nonumber\\
&&+\;\big(a_2-2b_2+\kappa/R\big)^2\Big\}+i\pi(\mu-\nu)+i\pi D(a_1a_2-b_1b_2)\nonumber\\
&&-\;\frac{2i\pi D}{5}\Big\{\big(2a_1-b_1-x_1-2\mu+\nu\big)\big(2a_2+b_2\big)-\big(a_1+2b_1-x_1/2-\mu-2\nu\big)\kappa/R\Big\} \nonumber\\
&&+\;\frac{2\pi}{5}\big(2A-B\big)\big(2k-l+1/2\big)-\frac{2i\pi}{5R}\big(k+2l+3/2\big)\kappa \nonumber\\
&&+\;\frac{i\pi}{R}\Big\{\big(j+1/2\big)\big(l+1/2\big)+Rx_1\big(l+1/2\big)-2\big(j+1/2\big)\big(k+1/2\big)\Big\}\bigg] \label{matrixel2}
\end{eqnarray}
if
\begin{equation}
0<2\re t'=\frac{2}{5}\Big(a_1+2b_1-x_1/2-\mu-2\nu\Big)<1 \label{ineqc}\\
\end{equation}
and otherwise zero. 

As in the previous cases, under the assumption $0<a_1,b_1<1$, we may use (\ref{ineqc})
to set $\mu=\nu=0$ and $x_1=\lfloor 2a_1 \rfloor$. Thus, we arrive at the following semi-classical 
approximation for our propagator
\begin{eqnarray}
\bra{b}\hat{B}_{1,-r}\ket{a} &=& \frac{1}{\sqrt{5}R^2}\msum{\kappa}\:\:\:\sum_{j,k=0}^{R-1}\:\:\:\sum_{l=0}^{2R-1}
\:\:\exp\bigg[-\frac{\pi D}{5}\Big\{\big(2a_1-b_1-\lfloor 2a_1\rfloor\big)^2+\big(a_2-2b_2+\kappa/R\big)^2 \nonumber\\
&&+\;i\big(3a_1a_2+3b_1b_2+4a_1b_2-4a_2b_1\big)-2i\kappa\big(a_1+2b_1-\lfloor 2a_1\rfloor/2\big)/R-2i\lfloor 2a_1\rfloor\big(2a_2+b2\big)\Big\} \nonumber \\
&&+\;\frac{2\pi}{5}\big(2a-b^*-\lfloor 2a_1\rfloor\big)\big(2k-l+1/2\big)-\frac{2i\pi}{5R}\big(k+2l+3/2\big)\kappa \nonumber\\
&&+\;\frac{i\pi}{R}\Big\{\big(j+1/2\big)\big(l+1/2\big)+R\lfloor 2a_1\rfloor\big(l+1/2\big)-2\big(j+1/2\big)\big(k+1/2\big)\Big\}\bigg]+o(1) \label{matrixel3}
\end{eqnarray}
Note that the summation index $\kappa$ remains unconstrained. Consequently, additional probabilistic 
`humps' emerge at locations other than those specified by the classical baker's map. 
In fact, in the region $0<b_1,b_2<1$, there will be $2R$ humps at the positions 
$(b_1,b_2)=(2a_1-\lfloor 2a_1 \rfloor,(a_2+\kappa/R)/2)$ where $-a_2R<\kappa<2R-a_2R$. 

Consider the simplest case $r=0$. Our semi-classical propagator is then
\begin{eqnarray}
\bra{b}\hat{B}_{1,0}\ket{a} &=& \sqrt{\frac{4}{5}}\sum_{\kappa=0}^1
\:\:\exp\bigg[-\frac{\pi D}{5}\Big\{\big(2a_1-b_1-\lfloor 2a_1\rfloor\big)^2+\big(a_2-2b_2+\kappa\big)^2+i\big(3a_1a_2+3b_1b_2+4a_1b_2-4a_2b_1\big) \nonumber\\
&&-\;2i\kappa\big(a_1+2b_1-\lfloor 2a_1\rfloor/2\big)-2i\lfloor 2a_1\rfloor\big(2a_2+b_2\big)\Big\}+i\pi\big(\lfloor 2a_1\rfloor-\kappa\big) \bigg] \nonumber\\
&&\cos\bigg[\frac{i\pi}{5}\big(2a-b^*-\lfloor 2a_1\rfloor\big)+\frac{\pi}{2}\big(\lfloor 2a_1\rfloor+1/2\big)-\frac{2\pi}{5}\kappa\bigg]+o(1)
\end{eqnarray}
defining two humps: one at a position specified by the classical baker's map,
$(b_1,b_2)=(2a_1-\lfloor 2a_1 \rfloor,(a_2+\lfloor 2a_1 \rfloor)/2)$, with an asymptotic size of
\begin{equation}
\big|\bra{b(a)}\hat{B}_{1,0}\ket{a}\big|^2 = \frac{4}{5}\cos^2\Big[\frac{\pi}{2}\big(a_2-1/2\big)\Big]+o(1)
\end{equation}
and another at $(b_1,b_2)=(2a_1-\lfloor 2a_1 \rfloor,(a_2+1-\lfloor 2a_1 \rfloor)/2)$ 
with the size
\begin{equation}
\big|\bra{b(a)}\hat{B}_{1,0}\ket{a}\big|^2 = \frac{4}{5}\sin^2\Big[\frac{\pi}{2}\big(a_2-1/2\big)\Big]+o(1).
\end{equation}
One interpretation of these equations could be that a {\it stochastic} mapping is implied in the classical limit: a point at $(a_1,a_2)$ has the 
probability $\cos^2[\pi(a_2-1/2)/2]$ of obeying the classical baker's map, and probability $\sin^2[\pi(a_2-1/2)/2]$ of 
ending up at $(2a_1-\lfloor 2a_1 \rfloor,(a_2+1-\lfloor 2a_1 \rfloor)/2)$. Notice that there is now a smooth transition 
of probabilities as one crosses the lines $a_2=0,1$.

Consider the size of our probabilistic humps in the general case. If we set 
$(b_1,b_2)=(2a_1-\lfloor 2a_1 \rfloor,(a_2+\kappa/R)/2)$, with $\kappa$ fixed, then
\begin{equation}
\bra{b(a)}\hat{B}_{1,-r}\ket{a} = \sqrt{\frac{4}{5}}\:\exp\bigg[\displaystyle\frac{i\pi D}{2R}\Big(2a_1\kappa+R\lfloor 2a_1\rfloor a_2-\lfloor 2a_1\rfloor\kappa\Big)\bigg]\Psi_{\kappa}(a)+o(1)
\end{equation}
where
\begin{equation}
\Psi_{\kappa}(a) = \frac{1}{2R^2}\:\:\:\sum_{j=0}^{R-1}\:\:
\frac{\cos^2\big[\pi Ra_2\big]}
{\sin\Big[\pi\Big(a_2-(j+1/2)/R\Big)\Big]\sin\Big[\displaystyle\frac{\pi}{2}\Big(a_2-\lfloor 2a_1\rfloor+\kappa/R-(j+1/2)/R\Big)\Big]}.
\end{equation}
\begin{figure}[t]
\includegraphics[scale=0.75]{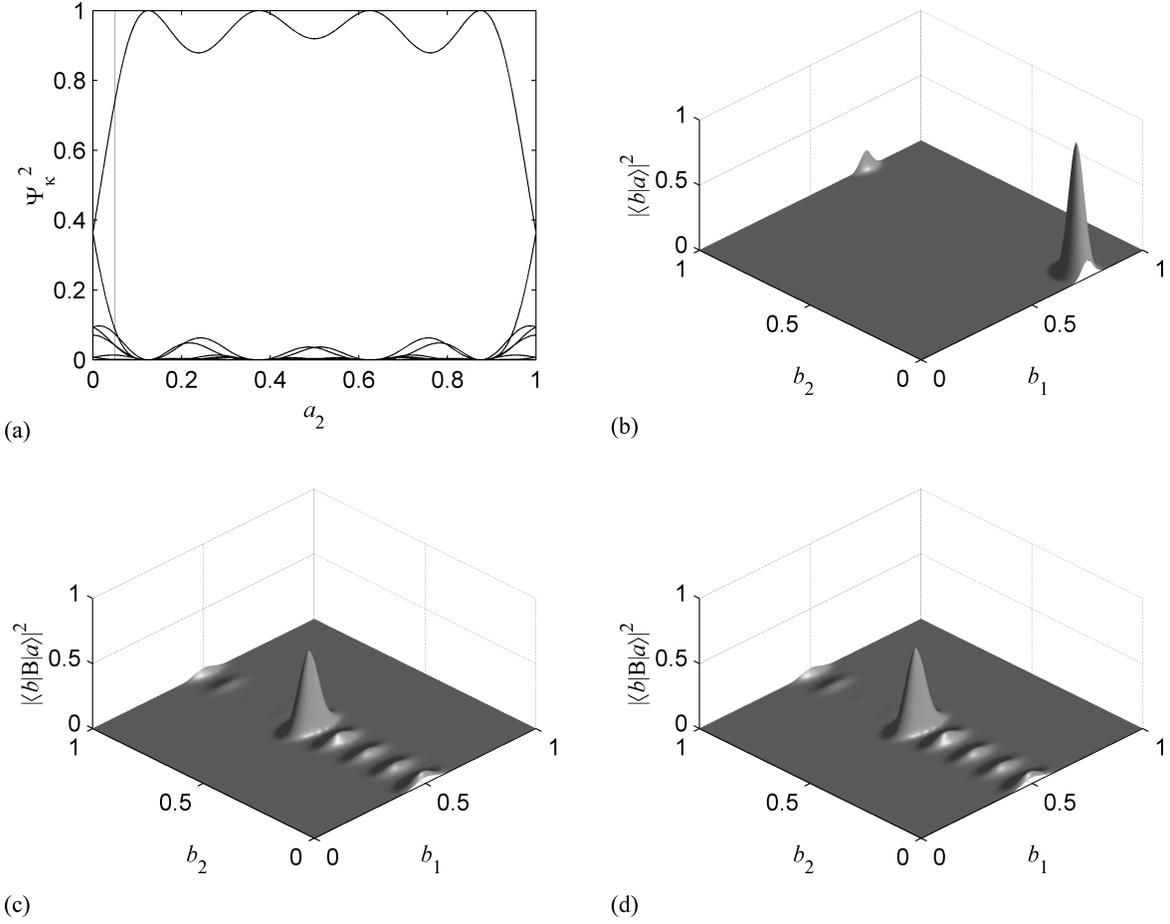}
\caption{(a) The probabilities $\Psi_{\kappa}^2$ when $r=2$. The Husimi functions of (b) the initial state $|a\rangle$
($a=3/4+i/20$), (c) its mapping $\hat{B}|a\rangle$, and (d) our semi-classical approximation (\ref{matrixel3}), when $r=2$ 
and $N=8$. The humps have the height $\frac{4}{5}\Psi_{\kappa}^2$ with $a_2=1/20$ (gray line in (a)).}
\label{humpic}
\end{figure}
The probability associated with each hump is given by $\Psi_{\kappa}^2$, with $\sum_{\kappa=0}^{2R-1}\Psi_{\kappa}^2=1$, 
and is plotted for $r=2$ in Fig. \ref{humpic}(a). The curve with the largest probabilities 
is that associated with the `correct' hump prescribed by the classical baker's map ($\kappa'=\lfloor 2a_1 \rfloor R$). 
It takes maximum values of unity whenever $a_2=(m+1/2)/R$ ($0<m<R-1$), and as the number of momentum 
bits ($r=\log_2R$) becomes large, 
$\Psi_{\kappa'}^2\rightarrow 1$ for all $0<a_2<1$. Hence, as expected, we are left with a single hump located at
the correct position. This agrees with the $0<\theta<1$ case. In figures \ref{humpic}(c) and (d), we have drawn the function 
$|\langle b|\hat{B}|a\rangle|^2$ and its semi-classical approximation (\ref{matrixel3}), respectively, when 
$a=3/4+i/20$, $r=2$ and $N=8$. For such a large dimension, $D=2^8$, our semi-classical approximation becomes 
almost identical to the exact matrix element. One may also view our semi-classical propagator as an approximation to the 
Husimi function for the mapped state $\hat{B}|a\rangle$.   

\section{Conclusion}

In this paper, we have derived the semi-classical form of the Schack-Caves quantum 
baker's maps, taking into consideration the different possible ways of obtaining the 
classical limit. We have shown that whenever the number of momentum bits becomes 
infinite in the limit $\hbar\rightarrow 0$, the quantum propagator in the coherent-state 
basis takes the Van Vleck form (\ref{vanvleck}). Therefore, we conclude that
the classical baker's transformation is restored for all such cases. In the case 
where the number of momentum bits is held constant, the classical limit is not that of 
the baker's transformation, but a stochastic variant. It may be possible one day to explore this 
discrepancy experimentally with the help of a quantum computer \cite{schack,brun}. As a final note, 
we remark that although our semi-classical formula (\ref{vanvleck}) espouses a certain familiarity, 
we should not be presumptuous about its generality. One should, in general, expect extra 
phases in the exponent (see Baranger {\it et al} \cite{Baranger}). This has already been found 
in a spherical geometry for the case of the kicked top \cite{Kus1993a}.

\begin{acknowledgments}
The authors would like to thank Carlton Caves for many helpful discussions and 
encouragement. This work was supported in part by the Office of Naval Research
Grant No. N00014-00-1-0578.
\end{acknowledgments}

\end{document}